\documentclass[aps,prd,nofootinbib,amsmath,amssymb,superscriptaddress,twocolumn,10pt]{revtex4}
\usepackage{graphicx}
\usepackage{dcolumn}
\usepackage{bm}
\usepackage{amssymb}
\usepackage{latexsym}
\usepackage{booktabs}
\usepackage{amsmath}
\usepackage{multirow}
\usepackage[colorlinks=true, linkcolor=red, citecolor=blue]{hyperref}

\newcommand{\be}{\begin{equation}}
\newcommand{\ee}{\end{equation}}
\newcommand{\bq}{\begin{eqnarray}}
\newcommand{\eq}{\end{eqnarray}}

\begin{document}

\title{Sterile neutrinos help reconcile the observational results of primordial gravitational waves from Planck and BICEP2}

\author{Jing-Fei Zhang}
\affiliation{Department of Physics, College of Sciences, Northeastern University, Shenyang
110004, China}
\author{Yun-He Li}
\affiliation{Department of Physics, College of Sciences, Northeastern University, Shenyang
110004, China}
\author{Xin Zhang\footnote{Corresponding author}}
\email{zhangxin@mail.neu.edu.cn} \affiliation{Department of Physics, College of Sciences,
Northeastern University, Shenyang 110004, China}
\affiliation{Center for High Energy Physics, Peking University, Beijing 100080, China}

\begin{abstract}
We show that involving a sterile neutrino species in the $\Lambda$CDM+$r$ model can help
relieve the tension about the tensor-to-scalar ratio $r$ between the Planck temperature data
and the BICEP2 B-mode polarization data. Such a model is called the $\Lambda$CDM+$r$+$\nu_s$
model in this paper. Compared to the $\Lambda$CDM+$r$ model, there are two extra parameters,
$N_{\rm eff}$ and $m_{\nu,{\rm sterile}}^{\rm eff}$, in the $\Lambda$CDM+$r$+$\nu_s$ model.
We show that in this model the tension between Planck and BICEP2 can be
greatly relieved at the cost of the increase of $n_s$.
However, comparing with the $\Lambda$CDM+$r$+$dn_s/d\ln k$ model that can significantly
reduce the tension between Planck and BICEP2 but also makes trouble to inflation due to the
large running of the spectral index of order $10^{-2}$ produced, the $\Lambda$CDM+$r$+$\nu_s$
model is much better for inflation. By including a sterile neutrino species in the standard cosmology,
besides the tension with BICEP2,
the other tensions of Planck with other astrophysical data, such as the $H_0$ direct measurement,
the Sunyaev-Zeldovich cluster counts, and the galaxy shear data, can all be significantly relieved.
So, this model seems to be an economical choice. Combining the Planck temperature data, the WMAP-9
polarization data, and the baryon acoustic oscillation data with all these astrophysical data
(including BICEP2), we find that
in the $\Lambda$CDM+$r$+$\nu_s$ model $n_s=0.999\pm 0.011$, $r=0.21^{+0.04}_{-0.05}$,
$N_{\rm eff}=3.95\pm 0.33$ and $m_{\nu,{\rm sterile}}^{\rm eff}=0.51^{+0.12}_{-0.13}$ eV.
Thus, our results prefer $\Delta N_{\rm eff}>0$ at the 2.7$\sigma$ level and a nonzero mass of sterile neutrino at the 3.9$\sigma$ level.
\end{abstract}

\pacs{95.36.+x, 98.80.Es, 98.80.-k} \maketitle

Detection of B-mode polarization of the cosmic microwave background (CMB)
was recently reported by the BICEP2 (Background Imaging of Cosmic Extragalactic Polarization)
Collaboration~\cite{bicep2}. The detected B modes might originate from the primordial gravitational waves
(PGWs) created by inflation during the very early moments of the universe.
If the BICEP2 result is confirmed by upcoming experiments, the frontiers of physics will be
pushed forward in an unprecedented way.

The BICEP2 Collaboration reported the fit result of the tensor-to-scalar ratio based on
the lensed-$\Lambda$CDM+$r$ model, $r=0.20^{+0.07}_{-0.05}$, from their observed
B-mode power spectrum data, with $r=0$ disfavored at the 7.0$\sigma$ level~\cite{bicep2}.
Subtracting the best available estimate for foreground dust slightly changes the likelihood but still results in
high significance of detection of $r$. 
However, the Planck Collaboration reported only a 95\% confidence level (CL) upper limit for the tensor-to-scalar ratio, $r<0.11$, from the fit to a combination
of Planck, South Pole Telescope (SPT) and Atacama Cosmology Telescope (ACT) temperature data, plus the Wilkinson Microwave Anisotropy Probe (WMAP)
9-year polarization data~\cite{planck}.
(Note that hereafter we use highL to denote the SPT+ACT data, and use WP to denote the WMAP-9 polarization data.)
Therefore, there is an apparent tension between Planck and BICEP2.

In order to reduce the tension, the BICEP2 Collaboration considered the case in which the running of the scalar spectral index, $dn_s/d\ln k$, is included.
For the Planck+WP+highL data combination, when the running is allowed, the fit results are~\cite{planck}:
$dn_s/d\ln k=-0.022\pm 0.010 $ (68\% CL) and $r<0.26$ (95\% CL), from which one can see that the tension between the previous TT measurements
and the current B-mode measurements is relieved.

However, it is well known that the usual slow-roll inflation models
cannot produce large running of the scalar spectral index; in these models $dn_s/d\ln k$ is typically of order $10^{-4}$.
In other words, the usual slow-roll inflation models cannot explain the large negative running of order $10^{-2}$ that is needed to reconcile the tension
between Planck and BICEP2.
A large running of the scalar spectral index is not good for inflation since the model must be contrived.
Therefore, in order to reduce the tension, more possibilities should be explored.

In fact, it has also been known that several astrophysical observations are inconsistent with the Planck temperature data.
For example, for the 6-parameter base $\Lambda$CDM model, from the Planck+WP+highL combination, it is found that the 68\% CL fit result of the Hubble constant
is $H_0=(67.3\pm 1.2)~{\rm km}~{\rm s}^{-1}~{\rm Mpc}^{-1}$~\cite{planck}, which is in tension with the direct measurement of the Hubble constant,
$H_0=(73.8\pm 2.4)~{\rm km}~{\rm s}^{-1}~{\rm Mpc}^{-1}$~\cite{h0}, at the 2.4$\sigma$ level.
Also, from the Planck temperature data, it seems that the standard cosmology predicts more clusters of galaxies than astrophysical observations see.
For the base $\Lambda$CDM model, the Planck+WP+highL data combination leads to $\sigma_8(\Omega_m/0.27)^{0.3}=0.87\pm 0.02$~\cite{planck},
while the counts of rich clusters of galaxies from an analysis of a sample of Planck thermal Sunyaev-Zeldovich (tSZ) clusters give
$\sigma_8(\Omega_m/0.27)^{0.3}=0.782\pm 0.010$~\cite{tsz}, thus there is a significant (4.0$\sigma$) discrepancy between them; the same data combination
leads to $\sigma_8(\Omega_m/0.27)^{0.46}=0.89\pm 0.03$~\cite{planck}, while the cosmic shear data of the weak lensing from the CFHTLenS survey give
$\sigma_8(\Omega_m/0.27)^{0.46}=0.774\pm 0.040$~\cite{wl}, thus there is a discrepancy at the 2.3$\sigma$ level.

One possible interpretation for these tensions
is that some sources of systematic errors in these
astrophysical measurements are not completely understood. However, there is an alternative explanation that the base $\Lambda$CDM model
is incorrect or should be extended.

Indeed, it is possible to alleviate the tensions between Planck and other astrophysical data by invoking new physics.
For example, the Planck's tension with the Hubble constant measurement might hint that dark energy
is not the cosmological constant~\cite{hde}.
In addition, recently, it was demonstrated~\cite{sterile1,sterile2,sterile3} that the Planck's tensions with the $H_0$ measurement, the counts of rich clusters, and
the cosmic shear measurements may hint the existence of sterile neutrinos.
If a sterile neutrino species is added, then clumping would occur more slowly due to its free-streaming damping, producing fewer clusters.
In other words, the sterile neutrinos can suppress the growth of structure, bringing the Planck data into better accordance with the counts of clusters.
Meanwhile, the sterile neutrinos can increase the early-time Hubble expansion rate and so change the acoustic scale,
leading the Planck fit result of $H_0$ into better agreement with the direct measurement.

Now that the sterile neutrinos can change the acoustic scale and the growth of structure, they may also
impact on the constraints on the scalar spectral index $n_s$ and the tensor-to-scalar ratio $r$.
In this work, we will explore this possibility in detail. We will show that the sterile neutrinos can also help
resolve the tension of $r$ between Planck and BICEP2.

In the base $\Lambda$CDM model, there are three active neutrino species, and so the effective number of relativistic species, $N_{\rm eff}$,
is equal to 3.046 (due to non-instantaneous decoupling corrections)~\cite{Mangano:2005cc}. 
Also, a minimal-mass normal hierarchy for the neutrino masses is assumed,
namely, only one massive eigenstate with $m_\nu=0.06$ eV.
In this paper, we consider a sterile neutrino model in which there exists one massive sterile neutrino in addition to
the two massless and one massive active neutrinos in the $\Lambda$CDM+$r$ model, and the active neutrino mass is kept fixed at 0.06 eV.
Since we add massive sterile neutrinos into the $\Lambda$CDM+$r$ model, the model considered in this paper is called $\Lambda$CDM+$r$+$\nu_s$ model.
Compared to the $\Lambda$CDM+$r$ model (with seven parameters), there
are two extra parameters, $N_{\rm eff}$ and $m_{\nu,{\rm sterile}}^{\rm eff}$.
In the case of a thermally-distributed sterile neutrino,
the effective sterile neutrino mass $m_{\nu,{\rm sterile}}^{\rm eff}$
is related to the true mass via $m_{\nu,{\rm sterile}}^{\rm eff}=(T_s/T_\nu)^3m_{\rm sterile}^{\rm thermal}=(\Delta N_{\rm eff})^{3/4}m_{\rm sterile}^{\rm thermal}$,
with $\Delta N_{\rm eff}=(T_s/T_\nu)^4=N_{\rm eff}-3.046$.
In the Dodelson-Widrow case the relation is  $m_{\nu,{\rm sterile}}^{\rm eff}=\chi_s m_{\rm sterile}^{\rm DW}$,
with $\Delta N_{\rm eff}=\chi_s$.

The possibility of the existence of light massive sterile neutrinos has been motivated to explain the anomalies of short baseline neutrino oscillation experiments,
such as the accelerator (LSND~\cite{lsnd} and MiniBooNE~\cite{miniboone}), reactor~\cite{reactor} and Gallium~\cite{gallium} anomalies.
It seems that the fully thermalized ($\Delta N_{\rm eff}\approx 1$) sterile neutrinos with eV-scale mass are needed to explain these results~\cite{Hannestad:2012ky,rev1,rev2}.
Cosmological observations may provide independent evidence in searching for sterile neutrinos.

In the following we shall use the current data to constrain the $\Lambda$CDM+$r$ model and the $\Lambda$CDM+$r$+$\nu_s$ model,
and see how the sterile neutrino impacts on the constraint results of $n_s$ and $r$,
as well as other observables, and 
if the evidence of existence of sterile neutrino can be found in the cosmological data currently available. 
In our calculations, the {\tt CosmoMC} code~\cite{cosmomc} is employed.

\begin{table}
\caption{Fit results for the $\Lambda$CDM+$r$ and $\Lambda$CDM+$r$+$\nu_s$ models.
Best fit values with $\pm 1\sigma$ errors are presented, but for the parameters that cannot be well constrained,
the $95\%$ upper limits are given.}
\label{table1}
\begin{tabular}{lcccc}
\hline\hline Model & $\Lambda$CDM+$r$ & &\multicolumn{2}{c}{$\Lambda$CDM+$r$+$\nu_s$} \\
             \cline{2-2}\cline{4-5}
             Data  & CMB+BAO  & & CMB+BAO &All \\ \hline
$100\Omega_{\rm{b}}h^2$           & $2.211\pm0.024$ &
                   & $2.250\pm0.030$&$2.282\pm0.028$ \\

$\Omega_{\rm{c}}h^2$           & $0.1186\pm0.0014$ &
                   & $0.1273^{+0.0054}_{-0.0061}$&$0.1271^{+0.0049}_{-0.0048}$ \\
$10^4\theta_{\rm{MC}}$           & $104.138\pm0.055$ &
                   & $104.050^{+0.076}_{-0.075}$&$104.050\pm0.070$ \\

$\tau$           & $0.091\pm0.013$ &
                   &$0.097^{+0.014}_{-0.015}$&$0.107^{+0.014}_{-0.016}$\\
$n_{\rm{s}}$           & $0.9632\pm0.0053$ &
                   & $0.985^{+0.012}_{-0.014}$&$0 .999\pm0.011$ \\
$r_{0.05}$           & $<0.13$ &
                   & $<0.19$&$0.191^{+0.036}_{-0.041}$ \\
$N_{\rm{eff}}$           & ... &
                   & $3.72^{+0.32}_{-0.40}$&$3.95\pm0.33$ \\
$m_{\nu,\,\rm{sterile}}^{\rm{eff}}  $           & ... &
                   &$<0.51$&$0.51^{+0.12}_{-0.13}$ \\
$\ln (10^{10}A_{\rm{s}})$           & $3.087\pm0.025$ &
                   & $3.12^{+0.030}_{-0.034}$&$3.140^{+0.031}_{-0.035}$ \\
\hline
$r_{0.002}$           & $<0.12$&
                   & $<0.20$&$0.207^{+0.041}_{-0.052}$ \\
$\Omega_{\Lambda}$           & $0.6952\pm0.0084$ &
                   &$0.6956\pm0.0093$&$0 .6952^{+0.0088}_{-0.0087}$ \\
$\Omega_{\rm{m}}$           & $0.3076\pm0.0084$ &
                   & $0.3044\pm0.0093$&$0.3048^{+0.0087}_{-0.0088}$ \\
$\sigma_8$           &$0.825\pm0.011$ &
                   & $0.812^{+0.038}_{-0.029}$&$0.759\pm0.012$\\
$H_0$           & $67.80^{+0.64}_{-0.63}$ &
                   & $70.8^{+1.7}_{-2.1}$&$71.5^{+1.4}_{-1.6}$ \\
$S_8^{\rm sz}$ & $0.857\pm0.015$ &
                   & $0.842^{+0.038}_{-0.029}$&$0.787\pm 0.009$ \\
$S_8^{\rm wl}$ & $0.876^{+0.019}_{-0.018}$ &
                   & $0.858^{+0.038}_{-0.030}$&$0.802\pm 0.010$ \\
\hline
$-2\ln\mathcal{L}_{\rm{max}}$ & 9809.018 && 9808.138&9867.634 \\
\hline
\end{tabular}
\end{table}

\begin{figure}[htbp]
\centering
\includegraphics[scale=0.9]{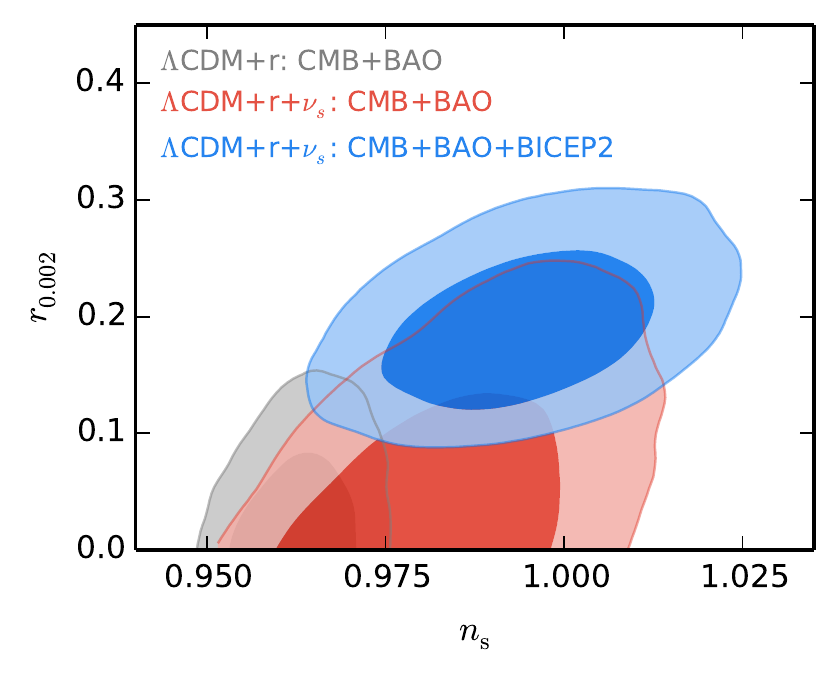} 
\caption[]{\small \label{fig1}Two-dimensional marginalized constraints (68\% and 95\% CL) on the scalar spectral index $n_s$ and the
tensor-to-scalar ratio $r_{0.002}$ for the $\Lambda$CDM+$r$ model and the $\Lambda$CDM+$r$+$\nu_s$ model. }
\end{figure}

First, we use the CMB+BAO data combination to constrain the models.
For convenience, hereafter we use CMB to denote the Planck+WP.
For the BAO data, we use the latest measurement
of the cosmic distance scale from the Data Release 11 galaxy sample of
the Baryon Oscillation Spectroscopic Survey (BOSS): 
$D_V(0.32)(r_{d,{\rm fid}}/r_d)=(1264\pm 25)$~Mpc and $D_V(0.57)(r_{d,{\rm fid}}/r_d)=(2056\pm 20)$~Mpc, with
$r_{d,{\rm fid}}=149.28$~Mpc~\cite{boss3}.

The constraint results in the $n_s$--$r_{0.002}$ plane are presented in Fig.~\ref{fig1}.
The grey contours are for the $\Lambda$CDM+$r$ model and the red contours are for the $\Lambda$CDM+$r$+$\nu_s$ model.
It is clear that in the $\Lambda$CDM+$r$+$\nu_s$ model the 95\% CL limit on $r$ is greatly relaxed, i.e., $r<0.20$,
but meanwhile the range of $n_s$ is also significantly enlarged and shifted towards the right.
So we conclude that in the $\Lambda$CDM+$r$+$\nu_s$ model the tension between Planck and BICEP2 can be
greatly relieved at the cost of the increase of $n_s$.
Detailed fit results for the $\Lambda$CDM+$r$ and $\Lambda$CDM+$r$+$\nu_s$ models from CMB+BAO
can be found in Table~\ref{table1}. Note that in our calculations the pivot scale is taken at $k_0=0.05$ Mpc$^{-1}$, but
in this paper $r$ always refers to $r_{0.002}$ (note that the fit values of $r_{0.05}$ are also given in Table~\ref{table1}).
Combining CMB+BAO and BICEP2 data gives $n_s=0.994^{+0.012}_{-0.013}$
and $r=0.19^{+0.04}_{-0.05}$ (95\% CL). The two-dimensional marginalized
posterior distribution between $n_s$ and $r$ in this case is given by the blue contours in Fig.~\ref{fig1}.
Note also that in our calculations with the B-mode data of BICEP2, the consistency relation for slow-roll
inflation, $n_t=-r/8$, is assumed.
With only the CMB+BAO data, however, $m_{\nu,{\rm sterile}}^{\rm eff}$ cannot be
tightly constrained, but only upper bound is given, $m_{\nu,{\rm sterile}}^{\rm eff}<0.51$~eV.
In order to precisely determine $N_{\rm eff}$ and $m_{\nu,{\rm sterile}}^{\rm eff}$, as analyzed in Refs.~\cite{sterile1,sterile2,sterile3},
other astrophysical data, such as $H_0$ measurement, SZ cluster data and lensing data,
should be considered.

\begin{figure}[htbp]
\centering
\includegraphics[scale=0.4]{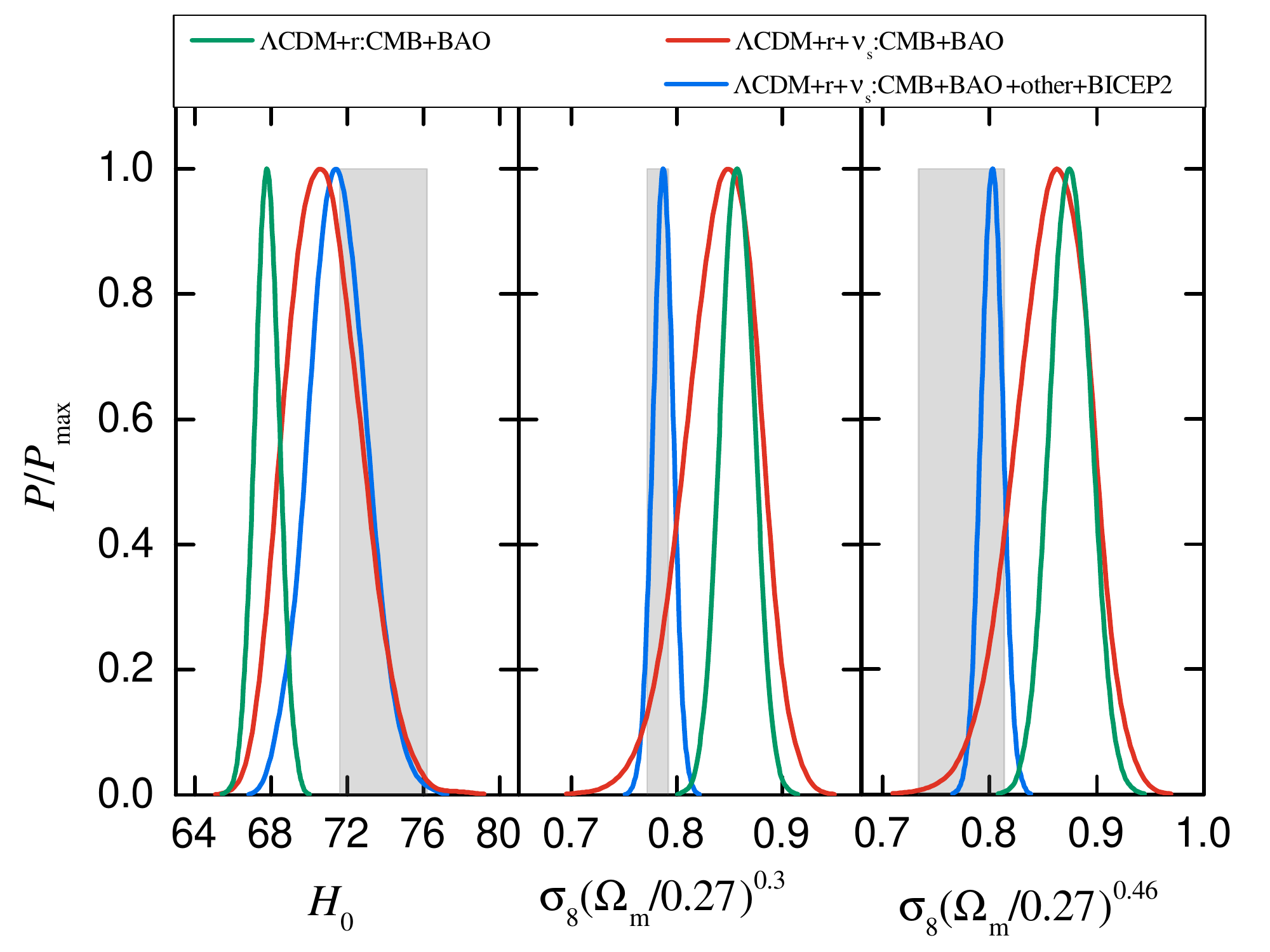}
\caption[]{\small \label{fig2}One-dimensional posterior distributions for $H_0$,
$\sigma_8(\Omega_m/0.27)^{0.3}$, and $\sigma_8(\Omega_m/0.27)^{0.46}$
in the $\Lambda$CDM+$r$ and $\Lambda$CDM+$r$+$\nu_s$ models.
Comparisons with the observational results are made. }
\end{figure}

\begin{figure}[htbp]
\centering
\includegraphics[scale=0.9]{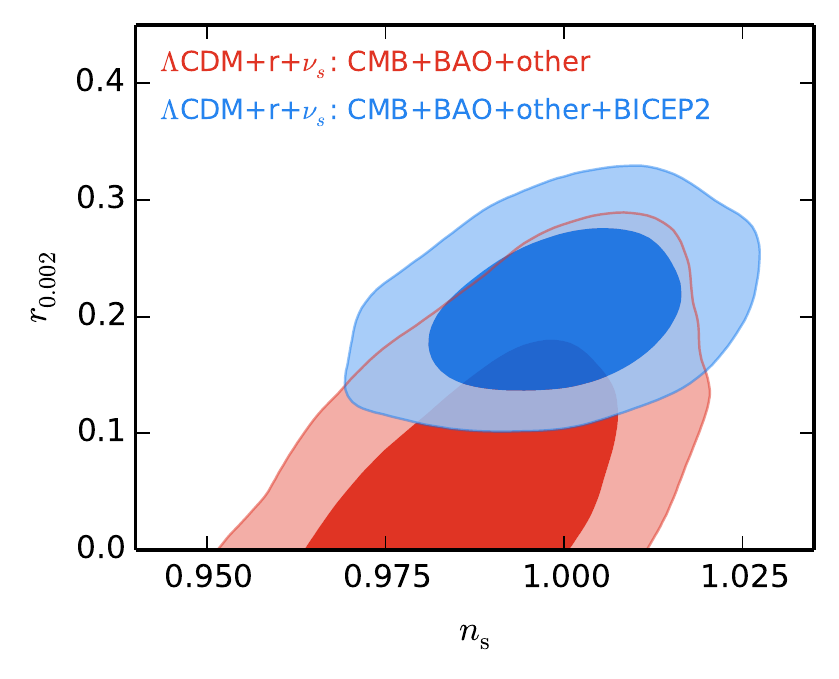}\\ 
\includegraphics[scale=0.9]{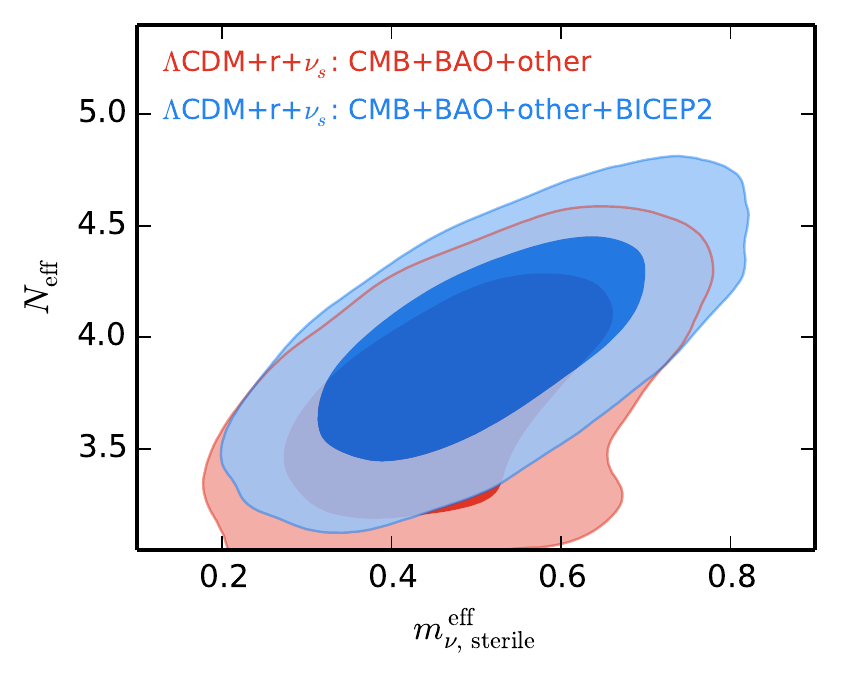}
\caption[]{\small \label{fig3}Two-dimensional joint, marginalized constraints (68\% and 95\% CL) on
the $\Lambda$CDM+$r$+$\nu_s$ model in the $n_s$--$r_{0.002}$ plane (upper) and in the
$m_{\nu,{\rm sterile}}^{\rm eff}$--$N_{\rm eff}$ plane (lower).}
\end{figure}

Next, we consider these astrophysical data. For the $H_0$ direct measurement, we use the HST
result $H_0=(73.8\pm 2.4)~{\rm km}~{\rm s}^{-1}~{\rm Mpc}^{-1}$~\cite{h0}.
For the SZ cluster counts, we use the Planck result $\sigma_8(\Omega_m/0.27)^{0.3}=0.782\pm 0.010$~\cite{tsz}.
For the lensing data, we use both the CMB lensing data $C_\ell^{\phi\phi}$ from Planck~\cite{cmblensing}
and the galaxy
lensing result $\sigma_8(\Omega_m/0.27)^{0.46}=0.774\pm 0.040$ from
CFHTLenS~\cite{wl}.
We shall first test the data consistency in the $\Lambda$CDM+$r$+$\nu_s$ model,
i.e., if the tensions between Planck and $H_0$, SZ cluster counts, and galaxy shear can be alleviated
at the same time in the $\Lambda$CDM+$r$+$\nu_s$ model so that the combination of these data sets
is appropriate.

The one-dimensional posterior distributions for $H_0$,
$\sigma_8(\Omega_m/0.27)^{0.3}$, and $\sigma_8(\Omega_m/0.27)^{0.46}$
in the $\Lambda$CDM+$r$ model and the $\Lambda$CDM+$r$+$\nu_s$ model are shown in Fig.~\ref{fig2},
where the green curves are for the $\Lambda$CDM+$r$ model constrained by CMB+BAO, and
the red and blue curves are for the $\Lambda$CDM+$r$+$\nu_s$ model constrained by
CMB+BAO and CMB+BAO+other+BICEP2, respectively.
Here, for convenience, we use ``other" to denote $H_0$+Lensing+SZ.
The observational results of $H_0$, SZ cluster counts, and galaxy shear are shown as the
grey bands in this figure. 
Comparing the green curves with the red curves, we find that the tensions of Planck with all these three
astrophysical observations are evidently alleviated.
The fit results of $H_0$, $\sigma_8(\Omega_m/0.27)^{0.3}$ and $\sigma_8(\Omega_m/0.27)^{0.46}$
for the three cases are given in Table~\ref{table1} (the last three rows), where the abbreviations $S_8^{\rm sz}$
and $S_8^{\rm wl}$ are used to denote $\sigma_8(\Omega_m/0.27)^{0.3}$ and $\sigma_8(\Omega_m/0.27)^{0.46}$,
respectively.
Under the constraints from CMB+BAO, considering the sterile neutrino improves the tension with $H_0$
from 2.4$\sigma$ to 1.0$\sigma$, the tension with SZ cluster counts from 4.3$\sigma$ to 2.0$\sigma$, and
the tension with cosmic shear from 2.3$\sigma$ to 1.7$\sigma$, respectively.
Thus, we find that the sterile neutrino not only can reconcile the $r$ results from
Planck and BICEP2, but also can simultaneously relieve almost all the tensions between Planck and
other observations.
Of course, residual tensions still exist, but this is rather natural because these astrophysical measurements
are sure to have some unknown systematic errors.

The constraint results in the $n_s$--$r_{0.002}$ and
$m_{\nu,{\rm sterile}}^{\rm eff}$--$N_{\rm eff}$ planes are shown in Fig.~\ref{fig3}.
In this figure, the red contours are for the CMB+BAO+other data combination,
and the blue contours are for the CMB+BAO+other+BICEP2 data combination.
We can see that in this case the $r$ tension between Planck and BICEP2 is further reduced, and
the parameters $N_{\rm eff}$ and $m_{\nu,{\rm sterile}}^{\rm eff}$ can be tightly constrained.
We find that for the $\Lambda$CDM+$r$+$\nu_s$ model, the CMB+BAO+other data combination gives
$n_s=0.991^{+0.015}_{-0.013}$, $r<0.23$ (95\% CL), $N_{\rm eff}=3.75^{+0.34}_{-0.37}$ and $m_{\nu,{\rm sterile}}^{\rm eff}=0.48^{+0.11}_{-0.13}$ eV,
and including the BICEP2 data modifies the constraint results to
$n_s=0.999\pm 0.011$, $r=0.21^{+0.04}_{-0.05}$, $N_{\rm eff}=3.95\pm 0.33$ and $m_{\nu,{\rm sterile}}^{\rm eff}=0.51^{+0.12}_{-0.13}$ eV.
We find that in the tightest constraints from the CMB+BAO+other+BICEP2 combination,
$\Delta N_{\rm eff}>0$ is at the 2.7$\sigma$ level and $m_{\nu,{\rm sterile}}^{\rm eff}>0$
is at the 3.9$\sigma$ level.
Our best-fit results, $\Delta N_{\rm eff}\approx 1$ and $m_{\rm sterile}^{\rm thermal}\approx m_{\nu,{\rm sterile}}^{\rm eff}\approx 0.5$ eV,
indicate a fully thermalized sterile neutrino with sub-eV mass. However, the short baseline neutrino oscillation experiments
prefer the mass of sterile neutrino at around 1 eV. The tension on the mass may deserve further investigations.
(For tension between short-baseline experiments and cosmology, see also Ref.~\cite{Mirizzi:2013kva}.)
It should also be pointed out that in the previous studies~\cite{sterile1,sterile2,sterile3} in which the $\Lambda$CDM+$\nu_s$ model is considered,
$\Delta N_{\rm eff}<1$ is preferred, 
but in this work we show that once the $\Lambda$CDM+$r$+$\nu_s$ model is considered, the full thermalization result ($\Delta N_{\rm eff}=1$)
compatible with the neutrino oscillation experiments can be obtained.

Finally, we wish to see how the observations of light elements abundances created during big bang nucleosynthesis (BBN) impact on the 
constraint results of sterile neutrino in the $\Lambda$CDM+$r$+$\nu_s$ model. 
Actually, a joint analysis including BBN observation has recently been made for the neutrino/dark radiation models without PGWs in Ref. \cite{neutrino14}. 
We shall follow Ref. \cite{neutrino14} to use the helium-4 and deuterium abundances to place constraints on the $\Lambda$CDM+$r$+$\nu_s$ model. 
The latest primordial $^4$He mass fraction measurement gives $Y_p=0.254\pm0.003$ \cite{he4}. For the primordial D fraction, we follow Ref. \cite{neutrino14} to 
consider two recent measurement values: $(D/H)_p=(2.87\pm0.22)\times 10^{-5}$ \cite{d1} and $(D/H)_p=(2.53\pm0.04)\times 10^{-5}$ \cite{d2}. 
So the two cases for BBN observation we consider are: (i) $Y_p$ from Ref. \cite{he4} + $(D/H)_p$ from Ref. \cite{d1} and (ii) $Y_p$ from Ref. \cite{he4} + $(D/H)_p$ from Ref. \cite{d2}. 
For the CMB+BAO+BBN combination, we obtain the constraint results of sterile neutrino: $N_{\rm eff}=3.75\pm0.20$ and $m_{\nu,{\rm sterile}}^{\rm eff}<0.45$ eV, for Case (i), 
and $N_{\rm eff}=3.38^{+0.14}_{-0.16}$ and $m_{\nu,{\rm sterile}}^{\rm eff}<0.59$ eV, for Case (ii). 
Furthermore, we consider the CMB+BAO+other+BBN combination, and we obtain the results: 
$N_{\rm eff}=3.78\pm0.19$ and $m_{\nu,{\rm sterile}}^{\rm eff}=0.46\pm0.10$ eV, for Case (i), 
and $N_{\rm eff}=3.45^{+0.14}_{-0.17}$ and $m_{\nu,{\rm sterile}}^{\rm eff}=0.43^{+0.10}_{-0.12}$ eV, for Case (ii).
We find that the consideration of BBN observation could tighten the constraints on $N_{\rm eff}$ and the constraint results are still consistent with the existence of sterile neutrino.

In this paper, in order to relieve the tension between Planck and BICEP2, we proposed to consider the light sterile neutrino in the model, i.e., the $\Lambda$CDM+$r$+$\nu_s$ model,
in which two extra parameters, $N_{\rm eff}$ and $m_{\nu,{\rm sterile}}^{\rm eff}$, are introduced.
In this model, not only the tension between Planck and BICEP2 is relieved, but also the tensions of Planck with other astrophysical observations are all alleviated at the same time. 
So in this work we actually made a comprehensive analysis for the $\Lambda$CDM+$r$+$\nu_s$ model. 
In fact, due to the galactic dust foreground contamination, the possibility that part or entire excess signal of PGWs could be explained by the dust emission 
cannot be excluded \cite{dust1,dust2,dust3}. But even so, we still provided a full analysis for the $\Lambda$CDM+$r$+$\nu_s$ model with the BICEP2 data optional, and we wish to stress that considering sterile neutrino could only enhance the upper limit of $r$ in the cases without adding BICEP2 data. Moreover, the mechanism of using sterile neutrino to reconcile various data sets in the $\Lambda$CDM+$r$+$\nu_s$ model has been 
discussed in detail and in depth in this work.


The inclusion of $N_{\rm eff}$ changes the early-time Hubble expansion rate and thus changes the acoustic scale, leading to the
change of the determination of $H_0$. Actually, $N_{\rm eff}$ is positively correlated with $H_0$.
Also, change of the acoustic scale leads to change of the determination of early-time parameters such as $n_s$.
In fact, $N_{\rm eff}$ is also positively correlated with $n_s$.
Increase of $N_{\rm eff}$ leads to increase of $n_s$, and thus the scalar powers on the large scales are suppressed,
leaving sufficient room for the contribution from the tensor powers.
Meanwhile, the increase of neutrino mass leads to the increase of the free-streaming damping, and so the growth of
structure is suppressed, producing fewer clusters.
Therefore, the inclusion of both $N_{\rm eff}$ and $m_{\nu,{\rm sterile}}^{\rm eff}$ can simultaneously relieve all the tensions of
Planck with other astrophysical observations.
We have also found from our analysis $\Delta N_{\rm eff}>0$ at the 2.7$\sigma$ level and a nonzero mass of sterile neutrino at the 3.9$\sigma$ level.
Detailed investigation on the implications of our results to inflation is the next step.

{\it Note added.}---After this paper was posted onto the arXiv (as arXiv:1403.7028), a few papers focusing on the similar subject also subsequently appeared on arXiv. 
A similar analysis was performed in Ref.~\cite{Dvorkin:2014lea}, where the Planck SZ cluster counts is replaced with the X-ray cluster result and the 
ACT/SPT temperature data are included. 
In Ref.~\cite{Archidiacono:2014apa}, an analysis of testing the consistency between the cosmological data (including BICEP2) and the neutrino oscillation data 
was performed. 
In Ref.~\cite{Zhang:2014nta}, we considered four neutrino cosmological models (i.e., $\Lambda$CDM+$r$+$\sum m_\nu$, 
$\Lambda$CDM+$r$+$N_{\rm eff}$, $\Lambda$CDM+$r$+$\sum m_\nu$+$N_{\rm eff}$, and $\Lambda$CDM+$r$+$N_{\rm eff}$+$m_{\nu,{\rm sterile}}^{\rm eff}$) 
and made a comparison for them. We showed that the former two models cannot get large $r$, and the third one can achieve large $r$ but makes additional assumption 
that massive active neutrinos coexist with some dark radiation. Thus, the result of Ref.~\cite{Zhang:2014nta} provide further support to this paper, explaining why we are most interested in 
the sterile neutrino case. In Ref.~\cite{Zhang:2014ifa}, further analyses were made for the cases in which the cosmological constant is replaced by the dynamical dark 
energy with constant $w$. In Ref.~\cite{Leistedt:2014sia}, it was shown that the sterile neutrino cosmological model 
is not favored over the standard $\Lambda$CDM model if the Bayesian evidence is used as a criterion for comparing models. Progress on this subject is still going on. 

\begin{acknowledgments}
We acknowledge the use of {\tt CosmoMC}. 
JFZ is supported by the Provincial Department of Education of
Liaoning under Grant No. L2012087.
XZ is supported by the National Natural Science Foundation of
China under Grant No. 11175042 and the Fundamental Research Funds for the 
Central Universities under Grant No. N120505003.
\end{acknowledgments}

\end{document}